\documentclass[aps,prb,preprint,showpacs]{revtex4}

\usepackage{amsmath}

\begin{document}

\title{New low frequency oscillations in quantum dusty plasmas}

\author{L.\ Stenflo, P.\ K.\ Shukla, and M.\ Marklund}

\affiliation{Department of Physics, Ume{\aa} University, SE--901 87 Ume{\aa},
  Sweden}
  
\begin{abstract}
  The existence of two new low-frequency electrostatic modes in quantum dusty plasmas
  is pointed out. These modes can be useful to diagnose charged dust impurities in 
  micro-electro-mechanical systems.  
\end{abstract}

\maketitle


Recently, there has been a great deal of interest in investigating high- and low-frequency
electrostatic modes \cite{haas-etal1,anderson-etal,haas-etal2,haas,garcia-etal,marklund} in an unmagnetized quantum plasma. For this purpose, the authors of Refs. \onlinecite{haas-etal1} and 
\onlinecite{haas-etal2,haas,garcia-etal} have used quantum transport models for the electrons and ions, and derived modified dispersion relations for Langmuir and ion-acoustic waves. However, in micro-electro-mechanical systems \cite{makowich-etal} one also encounters high-$Z$ charged dust impurities \cite{shukla-mamun}, which can affect the system. As a consequence, we expect that ultra-cold quantum plasmas can support new dust modes \cite{garcia-etal}. Examples of such quantum plasmas, as well as the range of validity of their existence, has been discussed recently in Ref.\ \onlinecite{manfredi}.

Let us consider a three component quantum plasma containing electrons, ions, and charged dust impurities with unperturbed number densities $n_{e0}$, $n_{i0}$, and $n_{d0}$, respectively. At equilibrium, we have $Z_in_{i0} = n_{e0} + \epsilon Z_dn_{d0}$, where $Z_i$ is the ion charge state,  $\epsilon = 1 \, (-1)$ for negatively (positively) charged dust particles, and $Z_d$ is the number of electrons/ions residing on the dust particulates. We assume that the latter are charged electrostatically. In the presence of an electrostatic field ${\bf E} = -\nabla\phi$, where $\phi$ is the electrostatic potential, the electron density perturbation $n_{e1}$ ($\ll n_{e0}$) in quantum plasmas is given by 
\begin{equation}\label{eq:electron}
  \nabla^2n_{e1} + \frac{4m_en_{e0}e}{\hbar^2}\phi = 0 ,
\end{equation} 
which is derived by combining the electron continuity and momentum equations assuming that $n_{e1} \ll (\hbar^2/T_{eF}m_e)\nabla^2n_{e1}$, and $\partial^2n_{e1}/\partial t^2 \ll (\hbar^2/4m_e^2)\nabla^4n_{e1}$. Here $m_e$ is the electron mass, $e$ is the magnitude of the electron charge, $\hbar$ is the Planck constant divided by $2\pi$, and $T_{eF}$ is the Fermi electron temperature. 

Let us now consider two cases. First, we assume that the charged dust impurities are immobile. Hence, the dynamics of the ions is governed by the continuity and momentum equations 
\begin{equation}\label{eq:ioncont}
  \frac{\partial n_{i1}}{\partial t} + n_{i0}\nabla\cdot{\bf v}_i = 0 ,
\end{equation}
and
\begin{equation}\label{eq:ionmom}
  m_i\frac{\partial{\bf v}_i}{\partial t} = -Z_ie\nabla\phi - \frac{T_{i}}{n_{i0}}\nabla n_{i1} + \frac{\hbar^2}{4m_in_{i0}}\nabla\left(\nabla^2 n_{i1}\right) ,
\end{equation}
respectively, where $n_{i1}$ ($\ll n_{i0}$) is the ion density perturbation, $m_i$ is the ion mass, and $T_{i}$ is the ion temperature. Equations (\ref{eq:electron}), (\ref{eq:ioncont}) and (\ref{eq:ionmom}) are closed by the Poisson equation
\begin{equation}\label{eq:Poisson}
  \nabla^2\phi = 4\pi e(n_{e1} - Z_in_{i1}) .
\end{equation}
We now combine (\ref{eq:electron})--(\ref{eq:Poisson}) to obtain 
\begin{equation}\label{eq:immobile}
  \left( \nabla^4 + K_q^4\right)\left( \frac{\partial^2}{\partial t^2} - V_{Ti}^2\nabla^2  + \frac{\hbar^2}{4m_i^2}\nabla^4 \right)\phi + \omega_{pi}^2\nabla^4\phi = 0 ,
\end{equation}
where $K_q = (16\pi n_{e0}/a_0)^{1/4}$ is the quantum wavenumber, $a_0 = \hbar^2/m_ee^2$ is the Bohr radius, $\omega_{pi} = (4\pi n_{i0}Z_i^2e^2/m_i)^{1/2}$ is the ion plasma frequency, and $V_{Ti} = (T_{i}/m_i)^{1/2}$ is the ion thermal speed.

Assuming that $\phi = \hat{\phi}\exp(i{\bf k}\cdot{\bf r} - i\omega t)$, where ${\bf k}$ is the wavevector and $\omega$ the frequency, we Fourier transform (\ref{eq:immobile}) to obtain
\begin{equation}\label{eq:immobile-disp}
  \omega^2 = k^2V_{Ti}^2 + \frac{\hbar^2k^4}{4m_i^2} + \frac{\omega_{pi}^2k^4}{k^4 + K_q^4} .
\end{equation} 
In quantum plasmas with cold ions and $K_q \gg k$, we thus have from (\ref{eq:immobile-disp})
\begin{equation}
  \omega \approx \frac{Z_i}{2}\left( \frac{n_{i0}}{n_{e0}} \right)^{1/2} \frac{\hbar k^2}{\sqrt{m_em_i}} ,  
\end{equation}
which is a new eigenfrequency for the case $n_{i0}/n_{e0} \gg m_e/m_i$. We note that in quantum plasmas with negatively (positively) charged dust impurities, we have $n_{i0}/n_{e0} > 1$ ($< 1$).

Second, we consider inertialess ions and mobile dust. Here, the ion density perturbation, for $V_{Ti}^{-2}\partial^2n_{i1}/\partial t^2 \ll \nabla^2n_{i1} \ll (m_iT_{i}/\hbar^2)n_{i1}$, is 
\begin{equation}\label{eq:iondensity}
  n_{i1} \approx -\frac{n_{i0}Z_ie\phi}{T_{i}} .
\end{equation}
The dust number density perturbation $n_{d1}$ ($\ll n_{d0}$) is obtained from \cite{shukla-mamun}
\begin{equation}\label{eq:mobiledust}
  \frac{\partial^2n_{d1}}{\partial t^2} + \frac{n_{d0}\epsilon Z_d e}{m_d}\nabla^2\phi = 0 ,
\end{equation}
where $m_d$ is the dust mass. Equations (\ref{eq:electron}), (\ref{eq:iondensity}) and (\ref{eq:mobiledust}) are then combined with
\begin{equation}
  \nabla^2\phi = 4\pi e(n_{e1} - Z_{i}n_{i1} + \epsilon Z_dn_{d1})
\end{equation}
to give
\begin{equation}\label{eq:mobile}
  \left[ \left( \nabla^2 - k_{Di}^2 \right)\nabla^2 + K_q^4 \right]\frac{\partial^2\phi}{\partial t^2} + \omega_{pd}^2\nabla^4\phi = 0 ,
\end{equation}
where $k_{Di} = (T_{i}/4\pi n_{i0}Z_i^2e^2)^{-1/2}$ is the ion Debye wavenumber and $\omega_{pd} = (4\pi n_{d0}Z_d^2e^2/m_d)^{1/2}$ is the dust plasma frequency. 

As before, we assume that $\phi = \hat{\phi}\exp(i{\bf k}\cdot{\bf r} - i\omega t)$ and obtain from (\ref{eq:mobile})
\begin{equation}\label{eq:mobile-disp}
  \omega^2 = \frac{\omega_{pd}^2k^4}{(k^2 + k_{Di}^2)k^2 + K_q^4} .
\end{equation}
In the long wavelength limit, viz.\ $k \ll k_{Di}$, the dispersion relation (\ref{eq:mobile-disp}) reduces to
\begin{equation}\label{eq:mobile-long}
  \omega = \frac{\omega_{pd}k}{(k_{Di}^2 + K_q^4/k^2)^{1/2}} . 
\end{equation}
It is interesting to note that for $K_q \gg (kk_{Di})^{1/2}$ we have from (\ref{eq:mobile-long})
\begin{equation}
    \omega \approx \frac{Z_d}{2}\left( \frac{n_{d0}}{n_{e0}} \right)^{1/2} \frac{\hbar k^2}{\sqrt{m_em_d}}  . 
\end{equation}
On the other hand, in the opposite limit $K_q \ll (kk_{Di})^{1/2}$, Eq.\ (\ref{eq:mobile-long}) gives the usual dust acoustic wave frequency \cite{rao-etal} $kC_D$, where $C_D = \omega_{pd}/k_{Di}$ is the dust acoustic speed.

To summarize, we have presented the dispersion properties of two new low-frequency electrostatic modes in an unmagnetized quantum dusty plasma. The new modes are associated with the inertia of the ions and dust impurities, as well as with a force caused by the quantum correlations of the electron number density fluctuations that balance the electrostatic force. The frequency spectra of these modes should be useful in investigations of the charge density of dust impurities in micro-electro-mechanical systems.  

\acknowledgments

This research was partially supported by the Swedish Research Council and the Deutsche Forschungsgemeinschaft (Bonn). 

 \newpage


\end{document}